# AN ENVELOPE –FUNCTION APPROACH FOR A ONE-DIMENSIONAL PHOTONIC CRYSTAL CONTAINING SINGLE NEGATIVE MATERIALS

**Munazza Zulfiqar Ali and Tariq Abdullah**

Centre of Excellence in Solid State Physics, Quaid-i-Azam Campus,
Punjab University, Lahore-54590, Pakistan

**Abstract:**

An envelope function approach is used to study the wave propagation in a one-dimensional Photonic Crystal containing single negative layers. This approach enables one to obtain the analytic expressions of the parameters of an equivalent effective medium. From these parameters, it is seen that the periodic structure appears to be an equivalent left-handed medium for the envelope function. For the nonlinear wave propagation the phenomenon of bistability and gap soliton formation near the edges of the zero-$\phi$ gap is discussed.

## 1. INTRODUCTION

The study of the wave propagation in Photonic Crystals has been a very active research area for more than a decade [1-5]. Conventional Photonic Crystals result from the periodic arrangement of different materials having positive values of electric permittivity and magnetic permeability is frequently taken to be one (right-handed medium). Gaps in such Photonic Crystals are produced due to Bragg reflections. Recently the experimental realization of metamaterials [6] has opened up the possibility of making unconventional Photonic Crystals. The metamaterials which can have the simultaneous negative values of electric permittivity and magnetic permeability [7] are known as double negative (DNG) materials or Left handed materials. Inclusion of left-handed materials in Photonic Crystals can result in many new and interesting phenomena for linear as well as for nonlinear wave propagation [8-11]. The metamaterials in which either of these quantities (electric permittivity or magnetic permeability) has a negative value are termed as Single Negative Materials (SNG). The SNG characterized by a negative permittivity is termed as epsilon-negative (ENG) and that having a negative permeability is termed as mu-negative (MNG). A pair of ENG and MNG is sometimes referred to as a conjugate pair [12]. Left-handed materials are not found in nature whereas there are many naturally occurring materials in which either permittivity or permeability is negative in some frequency range. Recently, wave propagation is being studied in Photonic Crystals containing alternate ENG and MNG layers [13-22]. The refractive index in a SNG material is imaginary which gives rise to evanescent modes in a bulk medium. In a one dimensional periodic structure containing alternate ENG and MNG layers, propagating modes can appear whose appearance can be explained in terms of a tight binding model in solid state physics. An interesting phenomenon that takes place in this structure is the appearance of a zero-$\phi$ gap. The Bloch phase is zero inside a zero-$\phi$ gap. It has been shown that the properties of the zero-$\phi$ gap are quite different form those of a Bragg gap [16-19] as it results from a different mechanism. Some applications have also been suggested to utilize these properties [17, 19].

Most of the investigations on the wave propagation in one-dimensional periodic structures containing SNG materials have been carried out by using layer-by-layer calculations. In that case the wave propagation is treated locally within each layer and every time we jump to the other layer we have to apply the boundary conditions at the interface between the two layers. It is possible to write an analytic solution for the wave propagation in a periodic medium provided certain specific conditions are fulfilled. Analytic approaches give an overall picture of the wave propagation and provide more insight into the nature of the processes taking place. Here we have applied the envelope function approach [23-25] which is based on the multiple scale method, to investigate the nature of the wave propagation in such a structure. The condition under which this approach can be applied in a structure containing alternate ENG and MNG layers is explained later on in this paper. This approach allows the periodic structure to be characterized by a few parameters avoiding the conventional piece-wise description of the structure. In this approach the electric field can be decomposed into a fast Bloch like component and a slow envelope function. As far as this slow envelope function is concerned, the periodic structure behaves as a homogenous slab of material for linear as well as nonlinear wave propagation. The fast Bloch-like components determine the local behavior of the electric field within the repeating units. These components determine the parameters of effective homogeneous medium and make the connection with the electric field in the surrounding medium. Recently, in a numerical study it has been shown [21] that a periodic structure with alternating ENG and MNG layers can be effectively treated as a left-handed medium. In the approach which we have used here, since the structure behaves as a homogenous medium for the envelope function, it becomes straightforward to establish this kind of similarity. Most of the studies reported so far on such structures have dealt with the case of linear wave propagation. The major advantage of the envelope function approach is that the case of nonlinear wave propagation can be treated effectively and with more insight when compared with numerical studies.

In section 2 we have outlined the mathematical model. The theoretical approach is discussed in detail in section 3. In sections 4 and 5, the approach is applied to a finite structure in the case of linear and nonlinear wave propagation respectively at the edges of the zero-$\phi$ gap. The concluding remarks have been presented in section 6.

## 2. The Mathematical Model.

Here we consider a one-dimensional structure of alternating layers 'a' and 'b' assumed to be ENG and MNG respectively. Since the SNG materials are inherently dispersive, the electric permittivity in layer 'a' and the magnetic permeability in layer 'b' are represented by a Drude model description where absorption has been neglected.

For layer 'a' which is assumed to be ENG, the linear values are defined as:

$$\varepsilon_a = 1 - \frac{\omega_{pe}^2}{\omega^2} \quad (1a)$$

$$\mu_a = 1 \quad (1b)$$

For layer 'b' which is assumed to be MNG, these values are:

$$\varepsilon_b = 1 \quad (2a)$$

$$\mu_b = 1 - \frac{\omega_{pm}^2}{\omega^2} \qquad (2b)$$

So the refractive indices in the two layers are given by:

$$n_a = \sqrt{\varepsilon_a \mu_a} \qquad (3a)$$
$$n_b = \sqrt{\varepsilon_b \mu_b} \qquad (3b)$$

The layer 'a' behaves as ENG layer in the frequency range $\omega < \omega_{pe}$ and the layer 'b' behaves as MNG layer in the frequency range $\omega < \omega_{pm}$. Here we have assumed that $\omega_{pe} = \omega_{pm}$, so the two layers behave as SNG layers in the same frequency range and the refractive index (although it is imaginary) has the same value in both layers. The widths of the two layers are represented by $d_a$ and $d_b$ which are assumed to be different. The zero-$\phi$ gap comes about due to the mismatch in local phase shift produced by the different widths of the two layers. The calculations done in this paper are based on the values:

$$\omega_{pe} = \omega_{pm} = 3.3 \times 10^{10} \, rad/\sec \qquad (4a)$$

$$d_a = 7 \times 10^{-3} \, m \qquad (4b)$$

$$d_b = 3 \times 10^{-3} \, m \qquad (4c)$$

$$d = d_a + d_b = d = 10^{-2} \, m \qquad (4d)$$

The dispersion relation for the normally incident wave in a periodic structure containing alternate layers of ENG and MNG materials can be written as [17]:

$$Cosqd = Cosh(k_a d_a)Cosh(k_b d_b) - \frac{1}{2}(\frac{\eta_a}{\eta_b} + \frac{\eta_b}{\eta_a})Sinh(k_a d_a)Sinh(k_b d_b) \qquad (5)$$

The local wave vectors in the two layers are imaginary whose absolute values are given by:
$$k_a = \sqrt{|\mu_a \varepsilon_a|} k_0 \qquad (6a)$$
$$k_b = \sqrt{|\mu_b \varepsilon_b|} k_0 \qquad (6b)$$

Where $k_0$ is the free space wave vector. The absolute values of the wave impedance in the two layers are:

$$\eta_a = \sqrt{|\frac{\mu_a}{\varepsilon_a}|} \qquad (7a)$$

$$\eta_b = \sqrt{|\frac{\mu_b}{\varepsilon_b}|} \qquad (7b)$$

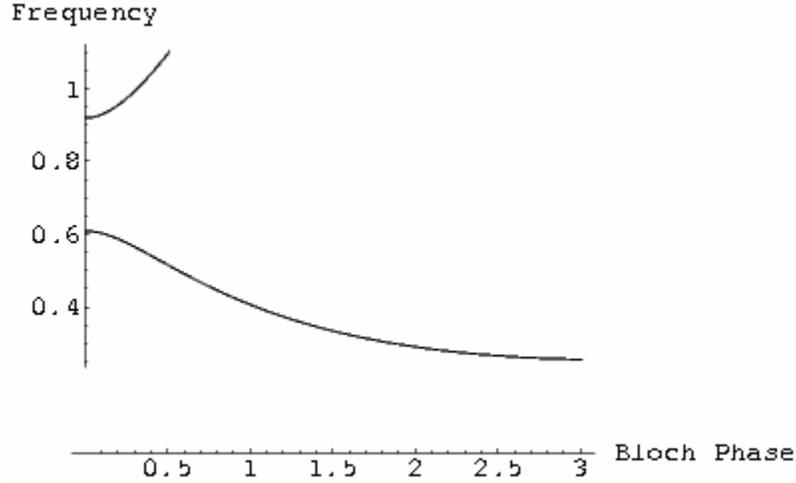

FIG.1. Dispersion curve for the structural parameters given in Eqs.(4a-d) in the frequency range in which the layers behave as SNG media. On the vertical axis, frequency is in dimensionless units i.e $W = \frac{\omega d}{c}$. The horizontal axis shows the Bloch phase '$qd$'

This dispersion relation is plotted in the Fig.1 for the frequency range in which the two layers behave as SNG media. At low frequencies, where $\eta_a \neq \eta_b$ $|Cosqd|>1$ corresponds to a gap and $|Cosqd|<1$ corresponds to propagating region. At a certain frequency the wave impedances in the two layers become identical i.e. $\eta_a = \eta_b$. If the widths of the two layers were taken to be same then we would have $k_a d_a = k_b d_a$ i.e. $|Cosqd|=1$ which corresponds to a resonant transmission i.e. the Bloch phase becomes zero. Since the widths of the two layers are taken to be different in the present case, there is a mismatch in the local phase shifts, i.e. $k_a d_a \neq k_b d_b$ at this wave impedance matching frequency which corresponds to $|Cosqd|>1$ and a gap is opened up which has been termed as the zero-φ gap as the Bloch phase is zero inside this gap. In the present arrangement the zero-φ gap is located in the frequency range *W=0.61- 0.92* (in dimensionless units) which corresponds to *18.2 - 27.5* in *GHz*.

## 3. The Theoretical Approach.

The wave equation for nonlinear wave propagation in this periodic structure can be written in CGS units as:

$$-c^2 \frac{\partial^2 E(x,t)}{\partial x^2} + \mu(x)\varepsilon(x)\frac{\partial^2 E}{\partial t^2} = 4\pi\mu(x)\chi^{(3)}(x)\frac{\partial^2}{\partial t^2}[E(x,t)]^3 \qquad (8)$$

We have made the assumption that the nonlinearity comes from the electric polarization being nonlinear in either ENG layer or MNG layer and the nonlinear polarization originates due to the third order susceptibility which is the most commonly occurring nonlinearity exhibited by both centrosymmetric and noncentrosymmetric media. The magnetic permeability is assumed to be linear in either layer. We want to solve Eq. (8) by using the envelope function approach [24] that was previously applied to a periodic structure of right-handed layers. Many new features emerge due to the inclusion of SNG layers. So instead of just presenting the results, we present here a somewhat detailed treatment of the problems. Before considering the nonlinear equation, let us initially consider the linear wave equation i.e $\chi^{(3)}(x) = 0$ so that we can write:

$$-c^2 \frac{\partial^2 E(x,t)}{\partial x^2} + \varepsilon(x)\mu(x)\frac{\partial^2 E(x,t)}{\partial t^2} = 0 \qquad (9)$$

Since $\varepsilon(x)\mu(x)$ is a periodic function having the same period as the period of the structure, the stationary solution of Eq. (9) can be written in terms of Bloch function:

$$E(x,t) = \varphi_m(x)e^{-i\omega_m t} + cc \qquad (10)$$

Provided the following eigen value equation is satisfied:

$$-c^2 \frac{\partial^2 \varphi_m(x)}{\partial x^2} = \varepsilon(x)\mu(x)\omega_m^2 \varphi_m(x) \qquad (11)$$

Since the product $\varepsilon(x)\mu(x)$ is negative in either layer, the propagating modes in this structure can appear if the orthogonality condition for the normalized Bloch functions can be written as:

$$\int \varphi_{m'}^* \varepsilon(x)\mu(x)\varphi_m = -\delta_{mm'} \qquad (12)$$

This different character of the Bloch function as compared to that found in a periodic structure consisting of right-handed layers results in many important features which we discuss as we proceed in this paper. An important consequence is that group velocity and phase velocity are opposite in direction as explained in the discussion that follows Eq.17. and Eq.18.

Using the multiple scale method by introducing
$x_i = s^i x$
$t_i = s^i t$ $\qquad i = 0,1,2,3....$ $\qquad, s << 1 \qquad (13)$

$$E(x_i, t_i) = se_1(x_i, t_i) + s^2 e_2(x_i, t_i) + s^3 e_3(x_i, t_i) + ......... \qquad (14)$$

Making these substitutions in Eq. (8) and equating the coefficient of equal powers of $s$ on both sides of the equation we get:

Coefficients of terms in $s$ give:

$$-c^2 \frac{\partial^2 e_1(x,t)}{\partial x_0^2} + \varepsilon(x_0)\mu(x_0)\frac{\partial^2 e_1(x,t)}{\partial t_0^2} = 0 \qquad (15)$$

Its solution can be written as:

$$e_1(x_0, x_1..; t_0, t_1..) = a(x_1, x_2..; t_1, t_2..)\varphi_m(x_0)e^{-i\omega_m t_0} \qquad (16)$$

Where the function of slow variables i.e. $a(x_1, x_2...)$ is termed as the envelope function and the Bloch function $\varphi_m$ varies on the scale of the fastest variable $x_0$.

Equating terms of equal powers of $s^2$ and following the same procedure as done in Ref. [24] but utilizing the different normalization condition in this case, we get:

$$\frac{\partial a}{\partial t_1} - \omega_m' \frac{\partial a}{\partial x_1} = 0 \qquad (17)$$

Where $\omega_m'$ is the group velocity of the envelope function. In a one-dimensional periodic structure containing right-handed materials [24] we have:

$$\frac{\partial a}{\partial t_1} + \omega_m' \frac{\partial a}{\partial x_1} = 0 \qquad (18)$$

The above two equations (Eq.17 and Eq.18) suggest that in a structure of alternating ENG and MNG layers the envelope function is moving with a group velocity which is opposite to the direction of the *x-axis* (negative group velocity) whereas in a structure of right handed layers, the envelope function is traveling in the direction of the x-axis (positive group velocity). Since the wave vector of the propagating mode lies along the positive direction of *x-axis* and the phase velocity is along the direction of the wave vector, so a positive value of group velocity refers to a situation when phase velocity and group velocity have the same direction and a negative group velocity refers to a situation when the phase velocity and the group velocity have opposite direction. One of the basic characteristics of a left-handed medium is that the phase velocity and group velocity have opposite directions. So in this respect, the structure considered here can be taken to be an equivalent left-handed medium.

Collecting the coefficient of $s^3$ and following the same procedure, a modified nonlinear Schrödinger equation is obtained:

$$i\frac{\partial a}{\partial t_2} - \frac{1}{2}\omega_m'' \frac{\partial^2 a}{\partial z_1^2} - \alpha_m |a|^2 a = 0 \qquad (19)$$

Where $z_1 = x_1 + \omega_m' t_1$ and $\omega_m''$ is the group velocity dispersion defined as:

$$\omega_m'' = \frac{c^2}{\omega_m}\langle m|m\rangle - \frac{(\omega_m')^2}{\omega_m} - \frac{4c^2}{\omega_m}\sum_l \frac{|\langle l|\Omega|m\rangle|^2}{\omega_m^2 - \omega_l^2} \qquad (20)$$

$$|m\rangle = \varphi_m \qquad (21)$$

$$\Omega = -ic\frac{\partial}{\partial x_0} \qquad (22)$$

And $\alpha_m$ is the effective nonlinear coefficient defined in this formalism as:

$$\alpha_m = 6\pi\omega_m L \int_0^L \chi^{(3)}(x)\mu(x)|\varphi_m(x)|^4 \, dx \qquad (23)$$

Where $L$ is the length over which the Bloch function is periodic. The variables $t_2$ and $z_1$ ( with subscripts removed) are identified as the slow variables and Eq. (19) becomes:

$$i\frac{\partial a}{\partial t} - \frac{1}{2}\omega_m'' \frac{\partial^2 a}{\partial z^2} - \alpha_m |a|^2 a = 0 \qquad (24)$$

The nonlinear Schrödinger equation admits a soliton solution for the envelope function when $\omega_m''$ and $\alpha_m$ have the same signs. However in a right-handed periodic medium, the signs of the second and the third term in Eq. (24) are positive, so positive values of $\omega_m''$ and $\alpha_m$ in a right-handed periodic medium correspond to negative values of $\omega_m''$ and $\alpha_m$ in the present case. This difference also results from the different character of the Bloch function found here (Eq.12).

**2a) Determination of Bloch Functions at the edges of zero-phi Gap.**

Up till now we have made no specific choice of the Bloch functions. Now we choose the Bloch functions to lie at the edges of the zero-ϕ gap where the group velocity vanishes. The Bloch functions at the edges of the gap must have real values since they correspond to the situation when no energy is being transported (zero group velocity). In the case of a periodic structure consisting of regular media, the Bloch functions at the edges of Brillouin zone are sine and cosine functions with a periodicity which is twice the periodicity of the lattice. In a periodic SNG structure, the Bloch function can not be written in the form of sine and cosine function because the slope of the electric field across the ENG-MNG interface becomes discontinuous. The zero-ϕ gap lies at the middle of the Brillouin zone where the Bloch phase is zero, so the periodicity of the Bloch functions at the edges of the zero-ϕ gap is equal to the periodicity of the lattice. So we can determine the Bloch functions at the edges of the zero-ϕ gap if we could determine the electric field in a unit cell at the frequency at which the lower edge and the upper edge of the zero-ϕ gap occur.

So if we consider a unit cell $0 < x < d$, then we can write:

$$E_a = C\exp(k_a x) + D\exp(-k_a x) \qquad 0 \leq x \leq d_a \qquad (25)$$

$$E_b = E\exp(k_b x) + F\exp(-k_b x) \qquad d_a \leq x \leq d \qquad (26)$$

Here we are considering a normally incident plane polarized wave so that the constants C,D,E,F can be determined by applying the continuity conditions of the tangential component of the electric and magnetic field across the interface of the ENG layer and MNG layer and the Bloch theorem in the usual way.

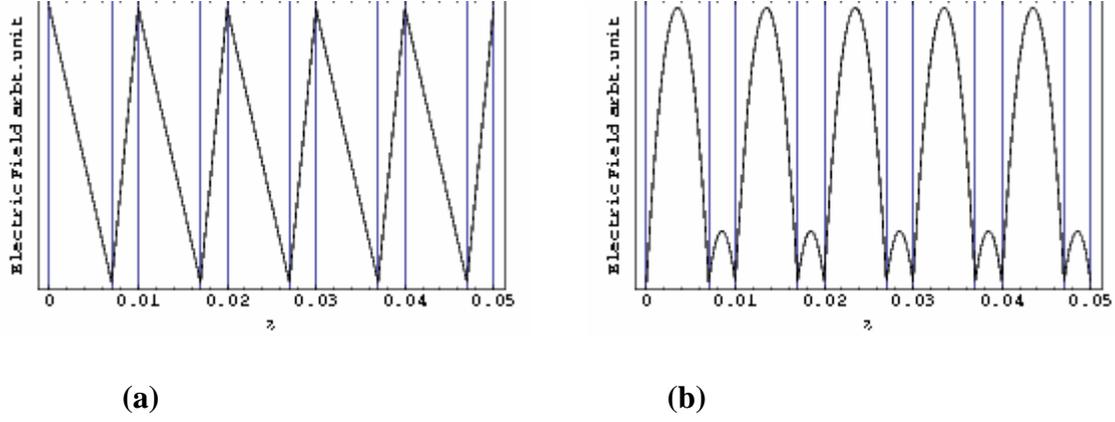

**(a)**  **(b)**

FIG.2.The electric field in five repeating units at frequency (a) $\omega = 18.2 \times 10^9 \, rad/s$ which corresponds to the lower edge of zero-$\phi$ gap (b) $\omega = 27.5 \times 10^9 \, rad/s$ The figure shows that the periodicity of the Bloch function is same as that of the repeating unit.

In Fig. 2 shows the Bloch function at the lower and the upper edge of the zero-$\phi$ gap. The slopes of the electric field are discontinuous across the ENG-MNG layers. This behavior resembles analytically with that found in a one dimensional structure containing alternate left-handed and right-handed layers because in both cases, there occurs a discontinuity in the slopes of the electric field at the interface of the two layers. Although in a LH-RH periodic structure propagating modes are present within the layers whereas in a SNG-MNG periodic structure decaying modes are present within the layers but these modes resemble one another over small range of distance.

**2b) Determination of the envelope function.**

Now we proceed to determine the envelope function. Assuming a harmonic time dependence for the envelope function i.e

$$a(x,t) = \psi(x) e^{-i\delta t} \qquad (27)$$

Here $\delta$ can be considered as the detuning i.e. $\delta = \omega - \omega_m$ and $\omega$ is the frequency of the incident radiation. So from Eq.(24) we get an equation of the form:

$$\frac{d^2\psi}{dx^2} - B^2 \psi + 2 \frac{B^2}{A^2} |\psi|^2 \psi = 0 \qquad (28)$$

Where :

$$A = \sqrt{\frac{2\delta}{\alpha_m}} \qquad (29a)$$

$$B = \sqrt{\frac{2\delta}{\omega_m''}} \qquad (29b)$$

The periodic structure thus appears to be a homogenous medium for the envelope function from Eq.(28). The solution of this equation determines the envelope function. The character of this effective medium depends upon parameters A and B. The parameters A and B must be real for the situations of main interest. In the present case, in order to have the real values of A and B, the sign of the detuning $\delta$ should be same as that of and $\alpha_m$ and $\omega_m''$ whereas in the case considered in Ref. [24] sign of $\delta$ must be opposite to those of $\alpha_m$ and $\omega_m''$ to get real values of A and B. Since the Bloch functions have been chosen to lie at the edges of the zero-ϕ gap, a positive detuning corresponds to a frequency which lies above the lower edge and a negative detuning corresponds to situation where the frequency lies below the upper gap edge inside the gap. Hence the lower edge of the zero- gap is associated with positive group velocity dispersion and the upper edge has negative group velocity dispersion. This is just the opposite behavior to that obtained in the case of Bragg gap. In order to ensure a real value of coefficient A, the positive value of $\alpha_m$ must shift the lower edge of zero- ϕ gap to higher value (positive detuning) and a negative value of $\alpha_m$ must shift the upper edge to a lower value. This again is the opposite behavior to that obtained in a Bragg gap. However, the sign of the effective nonlinear coefficient depends which of the two layers is taken to be nonlinear. If the electric permittivity of ENG layer is taken to be nonlinear, then a positive value of the $\chi^{(3)}(x)$ gives rise to a positive value of $\alpha_m$ and a negative value of $\chi^{(3)}(x)$ gives rise to a negative value of $\alpha_m$. If the electric permittivity of the MNG layer is taken to be nonlinear, then a positive value of $\chi^{(3)}(x)$ gives rise to a negative value of $\alpha_m$ and a negative value of $\chi^{(3)}(x)$ gives rise to a positive value of $\alpha_m$.

### 2c) Determination of the electric field.

Neglecting the third and the higher order terms and putting $s = 1$ in Eq.(14), the total electric field is written in terms of sum of two terms i.e.:

$$E(x) = \psi(x)\varphi_m(x) + \sum_l (Bd)(C_{l,m})(\frac{1}{B})\frac{d\psi}{dx}\varphi_l(x) \qquad (30)$$

$$C_{l,m} = \frac{2ic}{d}(\frac{\langle l|\Omega|m\rangle}{\omega_m^2 - \omega_l^2}) \qquad (31)$$

Where $\varphi_m(x)$ and $\varphi_l(x)$ are normalized Bloch functions and $C_{l,m}$ is the coupling constant between the state $l$ and $m$. The first term on the right hand side of Eq. (30) is called the principal term whereas the second term is called the companion term. The pre requisite to apply this approach is that the contrast in the refractive index of the two media should be small (zero in the limiting case) so that the summation in the second term on the right hand side of equation (30) reduces to only one term (Kogelnik approximation). In the case considered here, the zero-ϕ gap has emerged when there is no contrast in the refractive index of the two media. The mismatch in the local phase shifts is produced by taking the widths of the two layers to be different. So if the subscript $m$ refers to the state at the lower edge of the zero- ϕ gap, then the summation over $l$ includes only one term which is the state lying at the upper edge of the zero-ϕ gap. So, with known Bloch function and envelope function, Electric field can be determined at any point inside the medium by Eq.(30).

## 4. Linear Wave Propagation.

Initially we have investigated linear wave propagation in a finite structure. The purpose of considering the linear wave propagation is two-fold. We want to compare the results obtained by applying envelope function approach with those obtained by some exact method such as the transfer matrix approach in order to check the validity of this approach. Secondly, in order to simplify the calculations for the nonlinear wave propagation it can be shown [24] that the minima of the linear and nonlinear envelope function have the same value although they occur at different positions. So in order to determine the minima of the nonlinear envelope function, we have initially determined the minima of the linear envelope function.
Let us initially consider linear wave propagation in a finite structure of length $D_S$ whose transmitted end is assumed to be at $x=0$. The incident side lies at $x= -D_S$. The calculations done here assume that there is a ENG layer at $x=0$, so the first layer at $x= -D_S$ is assumed to be MNG .
In this case, the Eq. (26) reduces to

$$\frac{d^2\psi}{dx^2} - B^2\psi = 0 \qquad (32)$$

Its solution can be written as:

$$\psi = Pe^{Bx} + Qe^{-Bx} \qquad (33)$$

B is determined if the group velocity dispersion $\omega_m''$ and the detuning $\delta$ of the incident radiation is known by using Eq.(27b) where $\omega_m''$ can be determined in terms of the knowledge of the Bloch function of the structure by using Eq.20. So the envelope function can be determined if the coefficients P and Q are known. As usual these coefficients are determined by applying the boundary conditions on the tangential components of electric and magnetic fields at $x=0$ (which is taken to be the transmitted end of the structure), by assuming a plane wave solution outside the structure. Once P and Q are known, the electric field at any point $x$ inside the stack is found by the expression.

$$E(x) = (Pe^{Bx} + Qe^{-Bx})\varphi_m(x) + Bd(C_{l,m})(Pe^{Bx} - Qe^{-Bx})\varphi_l(x) \qquad (34)$$

In this paper, the incident frequency is taken to be 18.3 GHz which lies very close to lower edge of the zero-ϕ gap inside the gap. This is so because it can be shown [24] that the envelope function approach is valid under the condition when $\delta << \omega_l - \omega_m$. Fig 3 shows the field profile inside the structure at this frequency.

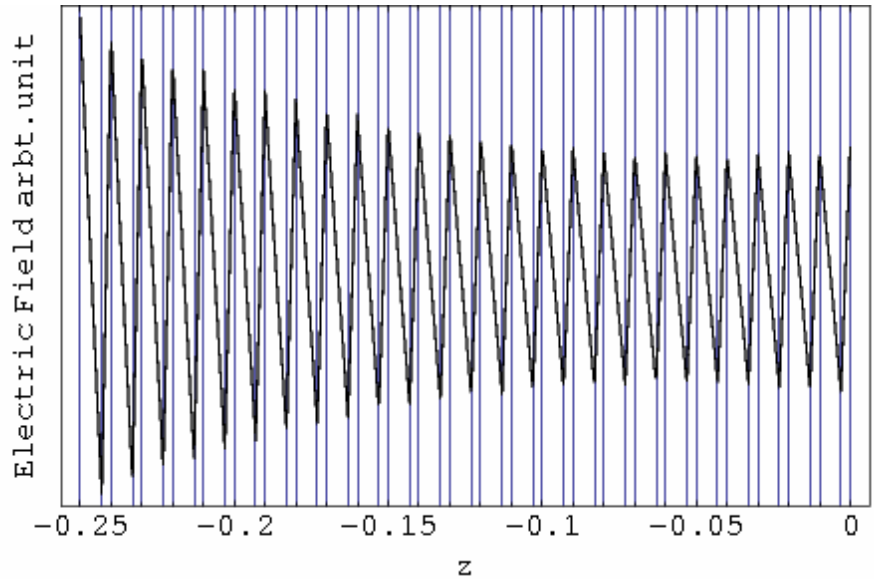

FIG.3. The electric field profile at frequency $\omega = 18.3 \times 10^9 \, rad/s$ which lies very close to the lower edge of the zero-$\phi$ gap inside the gap for linear wave propagation. The length of the structure is assumed to be 25 repeating units. The right side of the figure($x=0$) corresponds to the transmitted end whereas the left side of the figure corresponds to the incident side.

The transmission coefficient (T) is determined by applying the boundary conditions at the incident side of the structure. Fig. 4 shows the transmission coefficient as a function of the number of periods.

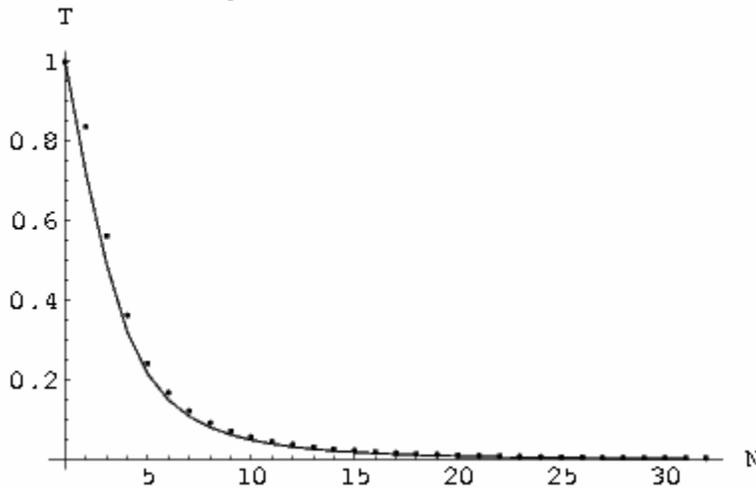

FIG.4. The plot of the transmission coefficient (T) versus the number of periods(N). The continuous line corresponds to the calculations done by the envelope function approach whereas the dots show the values of the transmission coefficient calculated by using the transfer matrix approach.

Using the envelope function approach, the transmission coefficient can be calculated as a continuous function of the length of the structure. Dots represent the transmission coefficient calculated by using the transfer matrix approach as a function of discrete number of periods. The figure shows that the calculations done by the envelope function approach are in good agreement with those carried out by using the transfer matrix approach.

## 4. Nonlinear Wave Propagation.

Recently the nonlinear response of a defect incorporated in a one-dimensional photonic crystal containing alternate MNG and ENG layer has been discussed [22]. We present here a more general case in which the overall nonlinear response of the structure (without any defect) is studied. As mentioned before, as far as the envelope function is concerned, the periodic structure is behaving as a homogenous medium. So if we introduce

$$\psi(x) = \xi(x)e^{i\varphi(x)} \quad (35)$$

In Eq.(26) and solve the equation for $I(x)$ where

$$I(x) = \xi^2(x) \quad (36)$$

Then I(x) can be written in terms of Jacobi elliptic function

$$I(x) = I_+ - (I_+ - I_m)cd^2[\sqrt{I_+ - I_-}\frac{B}{A}x \mid \kappa] \quad (37)$$

Where

$$I_+ = \frac{1}{2}[A^2 - I_m + [(A^2 - I_m)^2 + \frac{4W^2A^2}{I_m B^2}]^{1/2}] \quad (38)$$

$$I_- = \frac{1}{2}[A^2 - I_m - [(A^2 - I_m)^2 + \frac{4W^2A^2}{I_m B^2}]^{1/2}] \quad (39)$$

$$\kappa = [\frac{I_+ - I_m}{I_+ - I_-}]^{1/2} \quad (40)$$

*W* is the energy flux through the system.

The above parameters can be determined as a function of *W* if the value of $I_m$ can be found. As mentioned before $I_m$ can be determined by finding the minima of the linear envelope function whose modulus squared is taken to be equal to $I_m$. Then the envelope function for the nonlinear wave propagation is determined completely. The electric field at any point in the structure can be found by using Eq. (30) and hence the transmission coefficient is determined in the usual manner.

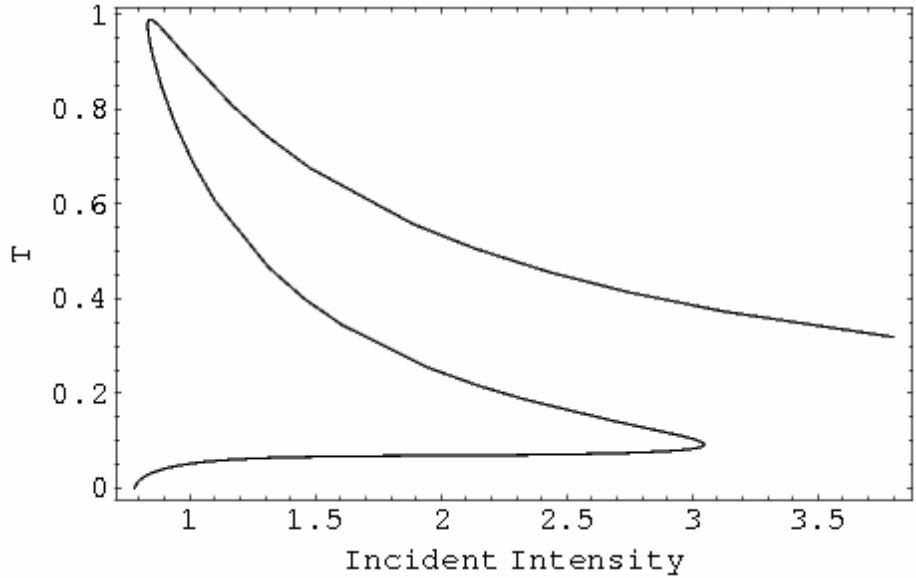

FIG.5. The plot of transmission coefficient (T) versus the incident intensity (in units of $W/cm^2 \times 10^3$ )for the same structural parameters as discussed for the linear wave propagation at $\omega = 18.3 \times 10^9 \, rad/s$ which lies just above the lower edge of the zero-ϕ gap. The ENG layer is taken to be nonlinear by taking $\chi^{(3)} = 10^{-4} \, e.s.u$ The length of the structure is taken to be 25 repeating units.

Figure 5 shows the transmission coefficient (T) versus the incident field intensity. It can be seen that the switch up and switch down fields are quite far apart. The Fig. 6 shows the field profile inside the structure when the transmission coefficient becomes unity i.e. it is the formation of a gap soliton [26].

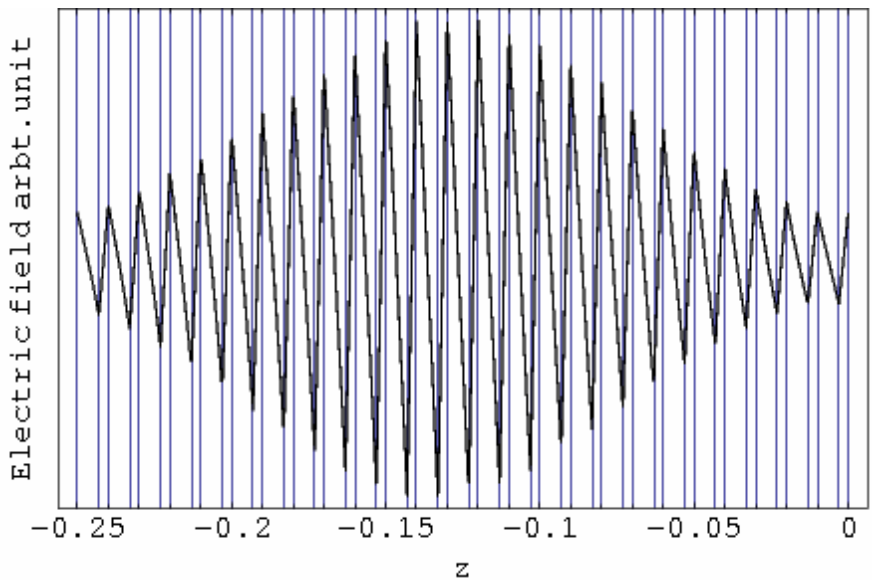

Fig.6 Electric field profile at the value of incident field intensity at which T=1, other parameters are same as those in Fig. (5).

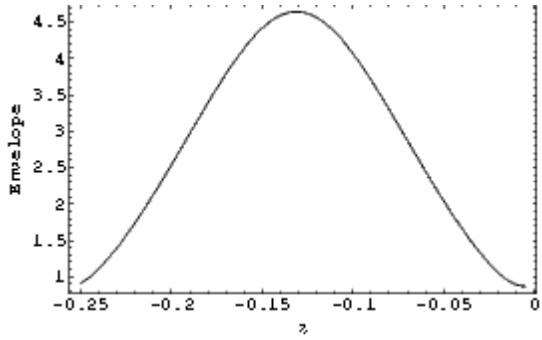 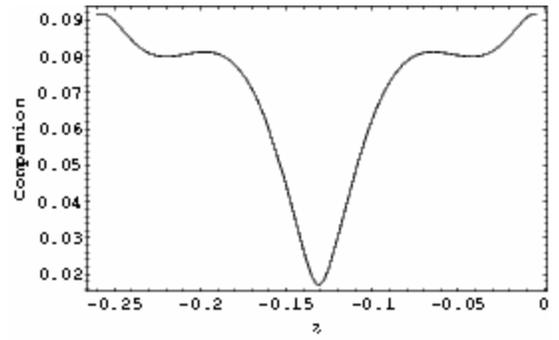

(a) (b)

FIG.7. The plot of the envelope function within the structure at transmission resonance(T=1) (a) associated with the principal term (b)associated with the companion term for the system parameters given in FIG.5. On the vertical axis the electric field is in arbitrary units.

Figure 7a shows the envelope function of the principal term associated with the gap soliton whereas Fig. 7b is the envelope function of the companion term at the same point. Both functions lie symmetrically in the structure. A comparison of the two plots shows that the contribution from the companion term is very small as compared to that of the principal term to the total electric field. It means that the reduction of the summation in the second term of Eq. (30) to only one term (Kogilnik approximation) is well justified in the present problem.

## 5. Conclusions.

We have investigated the wave propagation in a one-dimensional photonic crystal containing alternate MNG and ENG layers by using an envelope function approach. In this approach, the periodic structure appears to be an effective homogenous medium for the envelope function and the parameters of this effective medium such as group velocity, group velocity dispersion, and effective nonlinear coefficient can be calculated by using analytic expressions. So an overview of the processes taking place in the medium can be obtained.  We have compared the parameters of the effective medium in this case with that in a periodic structure composed of right-handed layers [24]. This comparison suggests that the periodic structure of ENG and MNG layers appears to be an equivalent left-handed medium for the envelope function. This character results from the different nature of the Bloch functions in this case as compared to that in a regular periodic medium. The nonlinear behavior of the structure at the edges of the zero-$\phi$ gap is found to be opposite to that at the edges of a Bragg gap. However, this behavior resembles the nonlinear behavior of a one dimensional structure consisting of alternate left-handed and right-handed layers [9, 11] where the nonlinearity in the right-handed layer gives rise to an effective nonlinear coefficient of the same sign but the nonlinearity in the left-handed layer reverses the sign of effective nonlinear coefficient.